\documentstyle[12pt,eqalign]{article}
\begin{document}
\begin{center} {\large\bf  Dependence of Electromagnetic Form Factors of
Hadrons on Light Cone Frames }\end{center}
\vskip 0.1in
\begin{center} {                      H. J. Weber   }\end{center}
\begin{center} { Institute of Nuclear and Particle Physics,  Department of
              Physics, University of Virginia, Charlottesville, Virginia
22901}\end{center}
\vskip 0.1in
\begin{center} {                      Xiaoming Xu  }\end{center}
\begin{center} { Institute of Nuclear and Particle Physics, Department of
Physics,
              University of Virginia, Charlottesville, Virginia 22901, and
 Shanghai Institute of Nuclear Research, Chinese Academy of
              Sciences, Shanghai 201800, China }\end{center}
\vskip 0.1in

\begin{abstract}
A constituent quark model is developed for an arbitrary light-cone
direction so that the light-front time is $x_{LF}^+=\omega \cdot x$ with a
constant lightlike four-vector $\omega$. Form factors are obtained from free
one-body electromagnetic current matrix elements. They are found to be
$\omega$-independent for spin-0 mesons, nucleons and the $\Lambda$ hyperon,
while there is $\omega $-dependence for spin-1 systems like the deuteron.
\end{abstract}
\vskip 0.3in
\begin{center} {\bf               I. Introduction  }\end{center}
\vskip 0.1in
\par
The light-front formalism of relativistic quantum mechanics has the attractive
feature that wave functions may be boosted kinematically, i.e. independent of
interactions. Since form factors necessarily involve boosted wave functions,
this aspect motivates many recent calculations in the light-front form of
relativistic few-body physics [1-8]. A basis of three-quark wave functions for
baryons has been constructed in light cone versions of the constituent quark
model (LCQM) to study the static properties
of the nucleon [1,2] and hyperons [3] and electromagnetic transition form
factors for $N \to N^*$ [2, 4] and $N \to\Delta$ processes[2,5]. The
three-quark wave function for a nucleon is the product of a
totally symmetric momentum wave function and a nonstatic spin wave function
which is an eigenfunction of the total angular momentum (squared) and its
projection on the light cone axis. The spin wave function can be represented
as a linear combination of products of matrix elements between valence quarks
[1,6] coupled by appropriate $\gamma$ (and isospin)-matrices to the
spin and isospin of the nucleon. Other nonstatic hadron wave functions are
constructed similarly. Direct coupling of spinors by Clebsch-Gordan
coefficients leads to equivalent spin-flavor wave functions [2].
Instant spinors are transformed into light-front helicity eigenstates by
Melosh transformations [9] which include the kinematic quark boosts. These
nonstatic spin invariants resemble the Ioffe currents for the nucleon that are
widely used in QCD sum rule techniques [10].
\par
In relativistic quantum mechanics the state vectors of a many-body system
transform as a unitary representation of the proper Poincar\'e group spanned
by ten generators, the four-momentum $P_{\mu}$ and the antisymmetric angular
momentum tensor $J_{\mu \nu}$ of the Lorentz group, which satisy the well known
commutation relations of the Poincar\'e algebra. Each form of relativistic
quantum mechanics is characterized by initial conditions on a hypersurface of
Minkowski space and a subgroup of kinematic (i.e. interaction independent)
generators in the stability group that map this hypersurface onto itself [11].
The standard instant form uses $t=constant$, where $t$ is the time, while the
front form is based on the null plane $ct+z=0$ [12]. The noninteracting front
form (or light cone) boosts form a
subgroup of the null plane's stability group that acts transitively in
momentum space. As a result the system's total and internal momentum
variables separate, whereas in the instant form no such separation is possible.
Thus, front form models are best developed in momentum space. In the instant
form the three-momentum $\vec{P}$ and angular momentum $\vec{J}$ are kinematic
and act transitively in coordinate space, while the three boost generators (of
rotationless Lorentz transformations along the spatial axes) are interaction
dependent so that, when the strong interaction is present, one simply cannot
use free boosts.
\par
The light-front form is obtained from the instant form in the
infinite momentum limit, and this amounts to the well-known change of momentum
variables $(p^+=p^{0}+p^z,p^-=p^{0}-p^z)$ on the light cone [13]. Since in
front form both transverse rotation generators become interaction dependent,
rotation invariance is more difficult to implement. In essence, light front
models include kinematic boosts at the expense of interaction dependent
angular momentum operators.
\par
Light front models obtained from quantum field theories usually involve
interacting spin operators. There is a unitary (but, as a rule, interaction
dependent) transformation to a representation where spins are free and state
vectors are specified on a null plane $\omega \cdot x=0$ with a lightlike (i.e.
$\omega^2=0$) four-vector $\omega^\mu$ [14,15,16]. If a theory is Lorentz
covariant its form factors are expected to be independent of $\omega$. In this
case the electromagnetic current operator in general has many-body
contributions [17]. When it is taken as  a sum of free one-body currents, as
in the impulse approximation for example, Lorentz covariance may be violated
which then shows up as explicit $\omega$-dependence in form factors.
In view of this situation it is worthwhile to test such frame dependence of
form factors. The conventional choice of the light cone direction is the
z-axis and the light cone time is $x^+=t+z$. Here we use instead an arbitrary,
but constant, lightlike four-vector $\omega^\mu$ to test form factors of the
constituent quark model in light front form (LCQM [1-8]) for their frame
dependence.
\par
In front form all form factors can be extracted from matrix
elements of $J^+$, the so-called 'good' current component [18]. When Lorentz
covariance and current conservation are imposed on the electromagnetic current
of a few-body system, current matrix elements can be parametrized by a minimal
number of Lorentz invariant form factors or multipole moments that are
characteristic of the composite system and its quantum numbers. Usually there
are more $J^+$ helicity matrix elements than form factors so that several
nontrivial constraints must be satisfied. The deuteron is a relevant example
for spin 1. Its electromagnetic properties are characterized by the charge,
magnetic and quadrupole form factors. Since there are four helicity matrix
elements, viz. $<\lambda'=1\mid J^+\mid \lambda=1>,
<\lambda'=1\mid J^+\mid \lambda=-1>, \newline
<\lambda'=1\mid J^+\mid \lambda=0>$
and $<\lambda'=0\mid J^+\mid \lambda=0>$, there is one angular condition
relating them. Often the longitudinal $<\lambda'=0\mid J^+\mid \lambda=0>$
matrix element is not used to extract the three form factors because the
helicity 0 states are not dominated by two-nucleon states [19]. The angular
condition is then also ignored; and there is $\omega$-dependence in the
deuteron case [14,15]. The electromagnetic $N\to \Delta$ transition is another
relevant case. For the rho-meson with J=1 such an angular condition has been
analyzed [20], but the J=0 and 1/2 cases are not considered as there is no
angular condition.
 When Lorentz invariance is violated, additional $\omega$-dependent current
components arise along with their form factors which are needed to fully
parametrize the current matrix elements that now obey fewer constraints. When
Lorentz invariance is restored by including two-body currents, these spurious
currents are cancelled and the remaining physical form factors are modified as
well. In other words, $\omega$-independence of form factors is just one step
toward Lorentz invariance but, by itself, it does not guarantee correct
form factors yet. In the case of $\omega$-dependence the unphysical currents
can at least be removed along with inconsistencies between current matrix
elements of different helicities.
\par
Since the electromagnetic current matrix elements are calculated with the free
current in quark models, the resulting form factors of hadrons may be frame (or
$\omega$-) dependent. In this paper we investigate this $\omega$-dependence
for spin-0 mesons and spin-$1 \over 2$ baryons in a
constituent quark model with boosts [1-8]. In Section II we set up light
front dynamics on the initial plane $\omega \cdot x=0$ by first defining the
light-front time and space coordinates and
other light-front quantities such as momentum variables,
Dirac $\gamma$ matrices, etc. Then the Dirac equation and free quark
spinors are presented in this frame. In Section III we calculate the
form factor of the $\pi$ meson and find that
it is independent of $\omega^\mu$. In Section IV the form factors of nucleon
and $\Lambda$ hyperon are calculated and found to be $\omega$-independent as
well. The final Section V contains the results and conclusion.
\vskip 0.2in
\begin{center} {\bf II. Light-Front Formalism  }\end{center}
\vskip 0.1in
In ordinary space-time coordinates, a light front hyperplane is described
by the invariant equation
$$
\omega' \cdot x =t- \vec {n'} \cdot \vec {x} =0            \eqno (2.1)
$$
where $\omega'^{\mu}=(1,\vec {n'})$ with $\vec {n'}$ the normal direction of
the light front hyperplane. The four-vector $\omega'$ satisfies $\omega'^2=0$
and its direction determines the light front surface. The
kinematics and dynamics constructed from the special light-front surface
$\omega'^\mu =(1, 0, 0, -1)$ is
the conventional light front choice. Here we develop a theory based on the
general light front surface with some constant $\omega'^{\mu}$. In coordinate
space we set up four orthogonal axes represented by the unit vectors
$s_{\mu}'^+$, $s_{\mu}'^-$, $s_{\mu}'^1$ and $s_{\mu}'^2$ with
$$
s_{\mu}'^+ \equiv \omega'_{\mu} =(1, -\vec n') =(1, -n'_x, -n'_y, -n'_z)
                      \eqno (2.2a)
$$
$$
s_{\mu}'^- =(1, \vec n') = (1, n'_x, n'_y, n'_z)            \eqno (2.2b)
$$
$$
s_{\mu}'^1 = (0,b_1,c_1, d_1)                \eqno (2.2c)
$$
$$
s_{\mu}'^2 = (0, b_2, c_2, d_2)                  \eqno (2.2d)
$$
so that $\vec {s'_1}$, $\vec {s'_2}$,$ \vec {n'}$ form the column vectors of a
rotation matrix
$$
R(\alpha, \beta, \gamma)$$
$$=\pmatrix {{cos \gamma cos \beta cos \alpha - sin \gamma sin \alpha}   &
         {cos \gamma cos \beta sin \alpha + sin \gamma cos \alpha}   &
         {-cos \gamma sin \beta}       \cr
        {-sin \gamma cos \beta cos \alpha - cos \gamma sin \alpha}    &
        {-sin \gamma cos \beta sin \alpha + cos \gamma cos \alpha}    &
        {sin \gamma sin \beta}         \cr
         {sin \beta cos \alpha}   &
         {sin \beta sin \alpha}     &    {cos \beta}   \cr}      \eqno (2.3)
$$
with the Euler angles $(\alpha, \beta, \gamma)$.
The four-dimensional frame spanned by the four axes is called a light-front
frame. A position vector denoted by $x^{\mu}=(t, x, y, z)$ in coordinate space
is represented in the new axes of the light-front frame as
$$
x_{LF}^+ =s_{\mu}'^+x^{\mu}=t-\vec {n'} \cdot \vec {x}          \eqno (2.4a)
$$
$$
x_{LF}^- = s_{\mu}'^-x^{\mu}=t+\vec {n'} \cdot \vec {x}        \eqno (2.4b)
$$
$$
x_{LF}^1 = s_{\mu}'^1x^{\mu}=b_1x +c_1y +d_1z           \eqno (2.4c)
$$
$$
x_{LF}^2 = s_{\mu}'^2x^{\mu}=b_2x +c_2y +d_2z           \eqno (2.4d)
$$
The subscript "LF" denotes the variables of the
light-front frame. From Eqs.(2.4a-2.4d), we define the inner, or scalar product
of $x_{LF}^{\mu}=(x_{LF}^+, x_{LF}^-,$
$x_{LF}^1, x_{LF}^2)$ with itself as
$$
x_{LF} \cdot x_{LF} =x_{LF\mu}x^{\mu}_{LF}
={1 \over 2}(x_{LF}^+ x_{LF}^- +x_{LF}^- x_{LF}^+)
-x_{LF}^1 x_{LF}^1 - x_{LF}^2 x_{LF}^2,              \eqno (2.5)
$$
so that $x_{LF} \cdot x_{LF}=x_{\mu}x^{\mu}$ in any frame. This is equivalent
to
$$
(\vec {n'} \cdot \vec {x})^2 + (b_1 x +c_1 y +d_1 z)^2 +(b_2 x+c_2 y +d_2 z)^2
=x^2 +y^2 +z^2,                   \eqno (2.6)
$$
which follows from the orthogonality of the rotation matrix R in Eq.(2.3).
\par
The axes in momentum space are then given by the row vectors $\vec s_1,
\vec s_2, \vec n$ of the rotation matrix R in Eq.(2.3) so that, e.g.,
$$
n_x=sin \beta cos \alpha                   \eqno (2.7a)
$$
$$
n_y=sin \beta sin \alpha                   \eqno (2.7b)
$$
$$
n_z=cos \beta.                      \eqno (2.7c)
$$
If
$$
s_{\mu}^+ \equiv \omega_{\mu} =(1, -\vec n) =(1, -n_x, -n_y, -n_z)
                      \eqno (2.2e)
$$
is given, then $\alpha$ and $\beta$ are determined. The
angle $\gamma$ is free and is used to fix the two axes $s_{\mu}^1$ and
$s_{\mu}^2$. The four-momentum $p^{\mu}=(p^0, p^1, p^2, p^3)$ is obtained by
projection onto the axes
$s_{\mu}^+, s_{\mu}^-,s_{\mu}^1$ and $s_{\mu}^2$
of the light-front frame as
$$
p^+_{LF} =s^+_{\mu} p^{\mu} =p^0 - \vec {n} \cdot \vec {p}      \eqno (2.8a)
$$
$$
P_{LF}^- =s_{\mu}^- p^{\mu} = p^0 + \vec {n} \cdot \vec {p}    \eqno (2.8b)
$$
$$
p_{LF}^1 = s_{\mu}^1 p^{\mu}                       \eqno (2.8c)
$$
$$
p_{LF}^2 = s_{\mu}^2 p^{\mu}                       \eqno (2.8d)
$$
Similarly, the matrix
$\gamma^{\mu}_{LF}=(\gamma^+_{LF}, \gamma^-_{LF}, \gamma^1_{LF},
\gamma^2_{LF})$ corresponding to
the Dirac matrix $\gamma^{\mu} =(\gamma^0, \gamma^1, \gamma^2, \gamma^3)$
in the light-front frame is defined as
$$
\gamma^+_{LF} =s_{\mu}^+\gamma^{\mu}=\gamma^0 - \vec {n} \cdot \vec {\gamma}
                 \eqno (2.9a)
$$
$$
\gamma^-_{LF} =s_{\mu}^- \gamma^{\mu} =\gamma^0 + \vec {n} \cdot \vec {\gamma}
                      \eqno (2.9b)
$$
$$
\gamma^1_{LF} =s_{\mu}^1 \gamma^{\mu}                \eqno (2.9c)
$$
$$
\gamma^2_{LF}=s_{\mu}^2 \gamma^{\mu}                  \eqno (2.9d)
$$
which gives $(\gamma^+_{LF})^2=(\gamma^-_{LF})^2=0$, $(\gamma^1_{LF})^2=
(\gamma^2_{LF})^2=-1$ and $\gamma^+_{LF} \gamma^-_{LF} +\gamma^-_{LF}
\gamma^+_{LF}=4$.
\par
We define the inner product of two four-vectors $a_{LF}^{\mu}=(a^+_{LF}$,
$a^-_{LF}$,
$a^1_{LF}$, $a^2_{LF})$ and $b_{LF}^{\mu} =(b_{LF}^+$, $b_{LF}^-$, $b_{LF}^1$,
$b_{LF}^2)$ as
$$
a_{LF} \cdot b_{LF} = {1 \over 2} (a_{LF}^+ b_{LF}^- +a_{LF}^- a_{LF}^+)
-a_{LF}^1 a_{LF}^1 -a_{LF}^2 a_{LF}^2               \eqno (2.10)
$$
In the light-front frame, the Dirac equation becomes
$$
(\gamma_{LF} \cdot p_{LF} -m)u_{LF}(p_{LF}) =0        \eqno (2.11)
$$
Its spin-up $u_{LF \uparrow}(p_{LF})$ and spin-down $u_{LF \downarrow}(p_{LF})$
solutions are helicity eigenstates in the infinite-momentum frame with
$p^+_{LF}\to \infty$,
$$
u_{LF \uparrow}(p_{LF})={1 \over \sqrt{2mp^+_{LF}}} (p^+_{LF} +\vec {\alpha}_
{LF \bot} \cdot \vec {p}_{LF \bot} + \beta m) \chi_{LF \uparrow}  \eqno (2.12a)
$$
$$u_{LF \downarrow}(p_{LF})={1 \over \sqrt{2mp^+_{LF}}} (p^+_{LF} +\vec
{\alpha}_{LF \bot} \cdot \vec {p}_{LF \bot} +\beta m ) \chi_{LF \downarrow},
                    \eqno (2.12b)
$$
in which $\vec {\alpha}_{LF \bot} = \gamma^0 \vec {\gamma}_{LF \bot}$,
$\beta=\gamma^0$ and
$$
\chi_{LF \uparrow} ={\sqrt{1-n_z} \over 2}
\pmatrix {1 \cr   -{n^R \over 1-n_z} \cr
1 \cr   -{n^R \over 1-n_z} \cr},             \eqno (2.13a)
$$
$$
\chi_{LF \downarrow} = {\sqrt{1-n_z} \over 2}
\pmatrix {{n^L \over 1-n_z} \cr      1 \cr
-{n^L \over 1-n_z} \cr    -1 \cr }             \eqno (2.13b)
$$
with $n^{R,L}=n_x \pm in_y$. When $n_x=n_y=0$ and $n_z=-1$, the $\chi$'s reduce
to the conventional light-cone spinors with
$$
\chi_{\uparrow}={1 \over \sqrt{2}}
\pmatrix {1 \cr   0 \cr  1 \cr  0 \cr} ,~~~~~~~~~~~~~
\chi_{\downarrow}={1 \over \sqrt{2}}
\pmatrix {0 \cr  1 \cr  0 \cr  -1 \cr}.                \eqno (2.14)
$$
The spinors satisfy the standard orthonormalization conditions
$$
\bar {u}_{LF \uparrow}(p_{LF}) u_{LF \uparrow}(p_{LF})
=\bar {u}_{LF \downarrow}(p_{LF}) u_{LF \downarrow}(p_{LF}) =1,   \eqno (2.15a)
$$
$$
\bar {u}_{LF \uparrow}(p_{LF}) u_{LF \downarrow}(p_{LF})
=\bar {u}_{LF \downarrow}(p_{LF}) u_{LF \uparrow}(p_{LF}) =0,     \eqno (2.15b)
$$
with
$$
\chi_{LF \uparrow}^{\dag } \chi_{LF \uparrow}=\chi_{LF \downarrow}^{\dag }
\chi_{LF \downarrow} =1,                \eqno (2.16a)
$$
$$
\chi_{LF \uparrow}^{\dag } \chi_{LF \downarrow}
=\chi_{LF \downarrow}^{\dag } \chi_{LF \uparrow} =0.         \eqno (2.16b)
$$
\par
The relevant projection operators are
$$
\Lambda_{LF+}={\gamma^-_{LF} \gamma^0  \over 2}, ~~~~~~~
\Lambda_{LF-}={\gamma^+_{LF} \gamma^0  \over 2}           \eqno (2.17)
$$
with the properties
$$
\Lambda^2_{LF+}=\Lambda_{LF+},~~~~~~\Lambda^2_{LF-}=\Lambda_{LF-},~~~~~~
\Lambda_{LF+}\Lambda_{LF-}=\Lambda_{LF-}\Lambda_{LF+}=0,
$$ ~~~~~~
$$
\Lambda_{LF+}+\Lambda_{LF-}=1.          \eqno (2.18)
$$
Then $\chi_{LF \uparrow}$ and $\chi_{LF \downarrow}$ are the eigenstates of
$\Lambda_{LF+}$ with the same eigenvalue 1,
$$
\Lambda_{LF+} \chi_{LF \uparrow} =\chi_{LF \uparrow},~~~~~~~
\Lambda_{LF+} \chi_{LF \downarrow} =\chi_{LF \downarrow},     \eqno (2.19a)
$$
$$
\Lambda_{LF-} \chi_{LF \uparrow}=\Lambda_{LF-} \chi_{LF \downarrow} =0.
             \eqno (2.19b)
$$
\par
Some basic matrix elements calculated from these spinors are shown in
Appendix 1 and Tables I, II. These matrix elements comprise
$\bar {u}_{\lambda_k} \Gamma u_{\lambda
_i}$ with $\Gamma=1, \gamma^+_{LF}, \gamma^-_{LF}, \gamma^{R,L}_{LF}=
\gamma^1_{LF} \pm i\gamma^2_{LF}$, $\gamma_5$, $\gamma_5 \gamma^+_{LF}$,
$\gamma_5 \gamma^-_{LF}$, $\gamma_5 \gamma^{R,L}_{LF}$,
where $\lambda_k$ and $\lambda_i$ are the helicities.
\par
We construct wave functions of hadrons in the light-front frame in the
constituent quark model as follows. First, we write down the spin-isospin
wave function in the hadron rest frame, then transform it into
the light-front spinors in Eq.(2.14) by Melosh transformations [9]; third,
rotate the wave function from the conventional light-front frame with
$n_x=n_y=0$ and $n_z=-1$ to that of Eq.(2.12) using the rotation matrix
$R(\alpha, \beta, \gamma)$. These wave functions are then used to calculate
electromagnetic form factors from the current $J^\mu_{LF}$ whose components
are defined as
$$
J^+_{LF}=s_{\mu}^+ J^{\mu},~~~~~J_{LF}^- =s_{\mu}^- J^{\mu},~~~~~
J_{LF}^1 =s_{\mu}^1 J^{\mu},~~~~~ J_{LF}^2 =s_{\mu}^2 J^{\mu}.
                 \eqno (2.20)
$$
\par
The Dirac matrices $\gamma^\mu_{LF}$ defined by Eq.(2.9) and
$\chi_{LF\uparrow, \downarrow}$ in Eq.(2.12) bring about the
$\omega$(or $\vec{n}$)-dependence in the calculation of the
electromagnetic current matrix elements in the following sections,
where also the cumbersome notation of subscript "LF" is dropped
for simplicity.

\vskip 0.2in
\begin{center} {\bf  III. $\pi$ Meson  }\end{center}
\vskip 0.1in
\par
To illustrate the calculation of the electromagnetic form factors of a spin-0
meson, we take the $\pi$ meson as an example.
Let $p_1$ and $p_2$ be the momenta of the valence quark and antiquark in the
meson, respectively. The + and $\bot$ components of the relative momentum and
total momentum are defined as
$$
q_3=x_2p_1-x_1p_2,~~~~~~~~~~P=p_1+p_2              \eqno (3.1)
$$
with Bjorken-Feynman variables $x_i (i=1, 2)$
$$
x_i={p_i^+ \over P^+},~~~~~x_1+x_2=1,~~~~~0 \leq x_i \leq 1.    \eqno (3.2)
$$
The momentum variables of the quark and antiquark in the
meson rest frame are
$$
\vec {k}_{1\bot}=\vec {p}_{1\bot}-x_1 \vec {P}_\bot,~~~~~
\vec {k}_{2\bot}=\vec {p}_{2\bot}-x_2 \vec {P}_\bot.       \eqno (3.3)
$$
A Gaussian momentum wave function is written in terms of the relative momentum
according to the Brodsky-Huang-Lepage prescription [21],
$$
\phi_o(x_i,q_3)=e^{-{M_2^2 \over 6\alpha^2}}          \eqno (3.4)
$$
with the hadronic size parameter $\alpha$ (and free mass operator $M_2$)
$$
M_2^2+\sum_{i=1}^2 {m_i^2 \over x_i}
=\sum_{i=1}^2 {\vec {k}_{i\bot}^2+m_i^2 \over x_i}
={\vec {q}_{3\bot}^2 +m^2 \over x_1x_2}         \eqno (3.5)
$$
for constituent quark mass $m_1=m_2=m$.
The light-cone wave function of the $\pi$ meson is [6, 8]
$$
\psi_{\pi}=N_{\pi}\phi_o \bar {u}_1 [P] \gamma_5 v_2,       \eqno (3.6)
$$
where $N_{\pi}$ is the normalization constant determined by the charge (form
factor) $F(0)=1$ for the $\pi^+$ meson. Here $u$ is the light-cone spinor of
Eq.(2.12) for the quark, $v_2=C\bar {u}_2^T$ with the charge-conjugation
operator $C$ for the antiquark, and the factor
$[P]={\gamma \cdot P +m_\pi \over 2m_\pi}$
with the $\pi$ meson mass $m_{\pi}$ originates from the Melosh rotation, while
$\gamma_5$ is characteristic of the spin 0 of the meson.
\par
In the light-front frame specified on the initial plane $\omega \cdot x=0$,
the electromagnetic current matrix elements of the free electromagnetic current
for a spin-0 meson consist of the usual convection current and an explicitly
$\omega$-dependent piece [14, 22],
$$
<\pi(P') \mid {J}^{\mu} \mid \pi(P)>=(P+P')^{\mu}F(q^2)
+\omega^\mu {(P+P')^2 \over 2\omega \cdot P} g_1(q^2),
                   \eqno (3.7)
$$
where $q=P'-P$, and $P'$ and $P$
are the initial and final momenta of the meson, respectively. To express
$F(q^2)$ and $g_1(q^2)$ in terms of the current matrix elements, we use
$$
\omega^+_{LF}=s^+_{\mu}\omega^{\mu}=0,~~~
\omega^-_{LF}=s^-_{\mu}\omega^{\mu}=2,~~~\omega^1_{LF}=s^1_\mu \omega^\mu =0,
{}~~~\omega^2_{LF}=s^2_{LF} \omega^{\mu}=0.       \eqno (3.8)
$$
The two form factors in the Drell-Yan frame $q^+=\omega \cdot q =0$
can then be expressed as
$$
F(q^2)={1 \over 2P^+}<\pi(P') \mid J^+ \mid \pi(P)>,          \eqno (3.9a)
$$
$$
g_1(q^2)={P^+ \over (P+P')^2}[<\pi(P') \mid J^- \mid \pi(P)> -
(P^-+P^{'-})F_1(q^2)].             \eqno (3.9b)
$$
In the impulse approximation the electromagnetic current matrix elements
are calculated from the triangle diagram alone in the Drell-Yan frame [22]
using the $\pi$-wave function of Eq.(3.6),
$$
<\pi(P') \mid J^+ \mid \pi(P)>
=\int d\Gamma \psi_{\pi}^{'\dag } (\sum_{j=1}^2 \bar {u}'_j e_j
{\gamma^+ \over x_j} u_j) \psi_{\pi}
$$
$$
\eqalign{
=N_{\pi}^2 \int d\Gamma \phi_o^{'*} \phi_o \{ e_1 Tr([P'][p_1']
{\gamma^+ \over x_1} [p_1][P][p_2])
+e_2 Tr([P'][p_1][P][p_2]{\gamma^+ \over x_2} [p_2']) \} },  \eqno (3.10)
$$
where $e_i$ is the electric charge and
$[p_i]={\gamma \cdot p_i +m_i \over 2m_i}$ the Melosh factor with $m_i$ the
mass and $p_i$ the momentum of the quark or antiquark, respectively.
\vskip 0.2in
\begin{center} {\bf IV. Electromagnetic Form Factors for Spin-$1 \over 2$
Baryons}\end{center}
\vskip 0.1in
\begin{center} {\bf A. Decomposition of Electromagnetic Current Matrix
Elements}
\end{center}
\vskip 0.1in
If $J^{\mu}$ is the full electromagnetic current, the current matrix
element of a baryon state consists of the standard Dirac and Pauli terms,
$$
<B(P')\lambda' \mid J^{\mu} \mid B(P)\lambda>
=\bar {u}_{\lambda'}(P') [\gamma^{\mu} F_1(q^2) +
{i \sigma^{\mu \nu} q_{\nu}  \over 2M} F_2 (q^2) ]
u_{\lambda} (\hat {P})             \eqno (4.1)
$$
with the baryon mass M.
\par
If the free quark currents are used, the possible $\omega$-dependence of the
electromagnetic current matrix elements between baryon states leads to three
additional components [14] so that Eq.(4.1) becomes
$$
<B(P')\lambda' \mid J^{\mu} \mid B(P)\lambda>
=\bar {u}_{\lambda'}(P') [\alpha_1 \gamma^{\mu}
+\alpha_2 {(P'+P)^{\mu} \over 2M} + \alpha_3 \omega^{\mu}
$$
$$
\eqalign{
{}~~~~~+ \alpha_4 \gamma \cdot \omega {(P'+P)^{\mu}
\over 2M} +\alpha_5 \gamma \cdot \omega \omega^{\mu} ]u_{\lambda}(P)}.
\eqno (4.2)
$$
The quantities $\alpha_1$ and $\alpha_2$ are
related to the form factors $F_1$ and $F_2$ by
$$
F_1+F_2=\alpha_1,           \eqno (4.3a)
$$
$$
F_2=-\alpha_2.              \eqno (4.3b)
$$
Taking the light-front wave function of a baryon in the LCQM to calculate the
matrix elements of the free current component $J^+$, we find by inspection that
$<B(P') \lambda' \mid J^+ \mid B(P) \lambda>$ is linear in
$P^+ +P'^+$, whereas the $\alpha_4$ term in Eq.(4.2) is proportional to
$(P^+ +P'^+)^2$. Thus, $\alpha_4=0$ in the LCQM.
 Direct algebraic calculation with the use of Tables I and II in the Drell-Yan
frame $q^+ =0$ with $\vec {P}_{\bot} =0$ in the initial state
yields the following expressions for the form factors:
$$
\alpha_1 ={M \over P^+} <B(P') \uparrow \mid J^+ \mid B(P)
\uparrow> - \alpha_2,                \eqno (4.4a)
$$
$$
\alpha_1={M \over P^+} <B(P') \downarrow \mid J^+ \mid B(P)
\downarrow > -\alpha_2,            \eqno (4.4b)
$$
$$
\alpha_2 = {2M^2d_1 \over P^+ n^Lq^L} <B(P') \uparrow \mid
J^+ \mid B(P) \downarrow>,          \eqno (4.4c)
$$
$$
\alpha_2 = -{2M^2d_1 \over P^+ n^R q^R} <B(P') \downarrow \mid
J^+ \mid B(P) \uparrow >,         \eqno (4.4d)
$$
$$
\alpha_3 = {M d_1 \over n^Lq^L } <B(P') \uparrow \mid J^- \mid
B(P) \downarrow> -\alpha_1 {M \over P^+} -\alpha_2 {\vec {q}^2_{
\bot} + 2M^2  \over 4MP^+},            \eqno (4.4e)
$$
$$
\alpha_5 ={M \over 2P^+} <B(P') \uparrow \mid J^- \mid B(P)
\uparrow> - {M^2d_1 \over P^+n^Lq^L} <B(P') \uparrow \mid
J^- \mid B(P) \downarrow>
$$
$$
+{M^2 \over 2(P^+)^2} \alpha_1.
                \eqno (4.4f)
$$
We note that the $J^-$ matrix elements in Eqs.(4.4e,f) being based on the
triangle diagram alone are incomplete, but are useful nonetheless for removing
the spurious $\omega$-dependent currents.
\vfill\eject\par
\begin{center} {\bf     B. Current Matrix Elements }\end{center}
\vskip 0.1in
  In the light-cone formalism all quarks are on their mass shell. Relative
momentum variables for the three valence quarks in a baryon may be defined
[1] in terms of the quark momenta $p_i$ by
$$
q_3 = (x_2 p_1 - x_1 p_2)/(x_1+x_2)                                \eqno (4.5a)
$$
$$
Q_3 = (x_1 + x_2) p_3 - x_3 (p_1 + p_2),                           \eqno (4.5b)
$$
which are independent of the total momentum P. These two equations are valid
for the + and
${\bot}$ components where $P^+=P_0+P_z$ and $\vec{P}_{\bot}=(P_x,P_y)$, etc.
The Bjorken-Feynman variables $x_j$ (j=1, 2, 3) are defined as
$$
x_j ={p_j^+ \over P^+},~~ \sum_{j=1}^3 x_j =1,~~ 0 \leq x_j \leq 1. \eqno (4.6)
$$
The ith quark momentum variable in the baryon rest frame is given by
$$
\vec {k}_{i\bot}=\vec {p}_{i\bot} -x_i \vec {P}_\bot.     \eqno (4.7)
$$
Let us denote the spin-isospin part of the light-front wave function by
$\chi_\lambda$. Then the wave function
of the baryon with helicity $\lambda$ is taken to be
$$
\psi_B(\lambda)=N_B\phi_o(x_3,q_3,Q_3)\chi_\lambda,       \eqno (4.8)
$$
where $N_B$ is a normalization constant fixed by the nonzero charge (form
factor at $q^2=0$) and the Gaussian momentum wave function
is
$$
\phi_o(x_i,q_3,Q_3)=e^{-{M_3^2 \over 6\alpha^2}}        \eqno (4.9)
$$
with light cone energy (and the free mass operator $M_3$)
$$
M_3^2+\sum_{i=1}^3 {m_i^2  \over x_i}
=\sum_{i=1}^3 {\vec {k}_{i\bot}^2 +m_i^2  \over x_i}
     =-q_3^2{1-x_3 \over x_1x_2}-{Q_3^2 \over x_3(1-x_3)}+
\sum_{i=1}^3 {m_i^2  \over x_i}.
$$
The light-front wave function is invariant under kinematic Poincar\'e
generators. It depends only on the relative momentum variables and the
longitudinal momentum fractions $x_i$ and the total mass eigenvalue, but not
on $P^{\mu}$.
\par
The matrix element of the free electromagnetic current in the Drell-Yan frame
[23] is obtained as
$$
<B \lambda' \mid J^+ \mid B \lambda>
=\int d\Gamma \psi^{'\dag}_B
(\sum_{i=1}^3 \bar{u}_i' e_i {\gamma^+ \over x_i} u_i) \psi_B
$$
$$
\eqalign{
= N^2_B \int d \Gamma \phi_o^*(x_i,q_3',Q_3')
\phi_o (x_i, q_3, Q_3) \chi_{\lambda'}^{\dag}
(\sum_{i=1}^3 \bar{u}_i' e_i {\gamma^+ \over x_i} u_i ) \chi_{\lambda}. }
                     \eqno (4.10)
$$
In the following two subsections the wave functions of the nucleon and
$\Lambda$
hyperon are employed to calculate electromagnetic form factors.
\vskip 0.1in
\begin{center} {\bf  C. Nucleon }\end{center}
\vskip 0.1in
The quark 1 and quark 2 are taken as up (down) quarks and quark 3 as a down
(up) quark for the proton (neutron) case. The nucleon wave function is [1]
$$
\psi_N=\pm N_N \phi_o \{ \bar {u}_2 [P] \gamma_5 C \bar {u}_3^T \cdot
\bar {u}_1 u_N - \bar {u}_3 [P] \gamma_5 C \bar {u}_1^T \cdot \bar {u}_2 u_N \}
                                          \eqno (4.11)
$$
with "+" for the proton and "-" for the neutron. The normalization constant
$N_N$ determined so that the Dirac form factor $F_1(0)=1$ for the proton. It
is a useful check of the symbolic and numerical codes to verify that $F_1(0)=0$
for the neutron. Upon relabeling the quarks so that
the quark 3 is always interacting with the electromagnetic field, the
matrix element of the free electromagnetic current becomes
$$
<p\lambda' \mid J^+ \mid p \lambda>
=N_N^2 \int d\Gamma \phi_o^{'*} \phi_o \{ 2e_u \bar {u}'_{\lambda'} [p'_3]
{\gamma^+ \over x_3} [p_3] u_\lambda \cdot Tr([P'][p_1][P][p_2])
$$
$$
+(e_u+e_d) \bar {u}'_{\lambda'} [p_2]u_\lambda \cdot Tr([P'][p'_3] {\gamma^+
\over x_3} [p_3][P][p_1])
$$
$$
+(e_u+e_d) \bar {u}'_{\lambda'} [p_1] u_\lambda \cdot Tr([P][p_3] {\gamma^+
\over x_3} [p'_3][P'][p_2])
$$
$$
+2e_u \bar {u}'_{\lambda'} [p'_3] {\gamma^+ \over x_3} [p_3][P][p_2][P'][p_1]
u_\lambda
$$
$$
+2e_u \bar {u}'_{\lambda'} [p_1][P][p_2][P'][p'_3] {\gamma^+ \over x_3} [p_3]
u_\lambda
$$
$$
\eqalign{
+2e_d \bar {u}'_{\lambda'} [p_1][P][p_3] {\gamma^+ \over x_3} [p'_3][P'][p_2]
u_\lambda   \}   \cr   }        \eqno (4.12)
$$
for the proton and
$$
<n\lambda' \mid J^+ \mid n\lambda> = <p\lambda' \mid J^+ \mid p\lambda>
(e_u \to e_d, e_d \to e_u)                   \eqno (4.13)
$$
for the neutron.
\vskip 0.1in
\begin{center} {\bf    D. $\Lambda$ Hyperon  }\end{center}
\vskip 0.1in
The up-down quark pair in the $\Lambda$ hyperon is coupled to zero spin and
isospin. The spin of $\Lambda$ hyperon is that of the $s$ quark.
Let $u$ be quark 1, $d$ quark 2  and $s$ quark 3. The light-front
wave function for the $\Lambda$ hyperon can then be written as
$$
\psi_{\Lambda} = N_{\Lambda} \phi_o \bar{u}_1[P] \gamma_5 v_2 \cdot \bar{u}_3
       u_{\Lambda},                                              \eqno (4.14)
$$
where $N_\Lambda$ is the normalization constant determined by $F_1(0)=1$ when
the charge $e_3=-1/3$ of the $s$ quark is replaced by +2/3. The matrix element
of the free electromagnetic current can then be written as
$$
<\Lambda \lambda' \mid J^+ \mid \Lambda \lambda>
=-N^2_\Lambda \int d \Gamma \phi_o^* \phi_o \{
 {2 \over 3} \bar{u}_{\lambda'}' [p_2]
 u_{\lambda} \cdot  Tr([P'][p_3'] {\gamma^+ \over x_3}
 [p_3][P][p_1])
$$
$$
 ~~~~ - {1 \over 3} \bar{u}_{\lambda'}' [p_1] u_{\lambda}
 \cdot  Tr([P'][p_2][P][p_3] {\gamma^+ \over x_3} [p_3'] )
$$
$$
\eqalign{
 ~~~~ -{1 \over 3} \bar{u}_{\lambda'}' [p_3'] {\gamma^+ \over x_3}
 [p_3] u_{\lambda} \cdot Tr([P'][p_1][P][p_2]) \} }
                                                        \eqno (4.15)
$$
where the labels now are so that quark 3 is always interacting with the
electromagnetic field.
\vskip 0.2in

\begin{center} {\bf       VI. Results and Discussion }\end{center}
\vskip 0.1in
When we construct the spin-isospin part of the wave functions of hadrons in the
constituent quark model in light-front form, the quarks and antiquarks are
assumed free. The Gaussian momentum wave function $\phi_0$ models the
confinement interaction, is Lorentz invariant and directly linked with the
successful spectroscopy of the nonrelativistic constituent quark model. Matrix
elements of free quark currents represent the lowest-order electromagnetic
Feynman diagram calculation including a confinement interaction in the bound
states.
\par
The instant form is usually specified by the time $t=0$ and the light-front
plane by $x^+=t+z$. When spin operators are taken as free, then a unitary
transformation to a more general initial null plane $\omega \cdot x=0$ becomes
unavoidable. This is no problem for a Lorentz covariant theory which will not
depend on the angles characterising this plane. The Poincar\'e group
realized on the initial plane $\omega \cdot x=0$ has seven kinematical
generators and three dynamical generators. The advantages of the conventional
light-front quantum field theory discussed in the introduction remain valid in
a field theory constructed from the initial plane $\omega \cdot x=0$. The
Hamiltonian $P^-$ among the dynamical Poincar\'e generators transforms the
initial plane $\omega \cdot x=0$ to another light-front plane. Such a dynamical
process evolves in the light-cone time $x_{LF}^+ = t- \vec {n} \cdot \vec x$
of Eq.2.4b. The two transverse space coordinates $x_{LF}^1$ and $x_{LF}^2$
are not completely specified by the $\vec n$, but depend also on the angle
$\gamma$ which is arbitrary. The most convenient choice is $\gamma=0$. We have
defined in Section II the momentum variables and Dirac matrices in terms of
these coordinates. The Dirac equation (2.11) has the same form in this
light-front frame.
\par
The Dirac matrices $\gamma^\mu_{LF}$ defined by Eq.(2.9) and
$\chi_{LF\uparrow, \downarrow}$ in Eq.(2.12) bring about the
$\omega$-dependence in the calculation of the
electromagnetic current matrix elements. In the $\pi$ meson case, the
$n^R$ and $n^L$ lead to $\omega$-independent current matrix elements because
${n^R n^L \over d_1^2}=1$ . The form factor $F(q^2)$
in Eq.(3.9a) stays $\omega$-independent even if the free quark current is used
in the calculations; it is physical and not influenced by the nonphysical form
factor $g_1(q^2)$. The wave function of each spin-0 meson has the same
structure as in Eq.(3.4) for the pion. Therefore we conclude that the physical
form factor of the spin-0 meson may be extracted from the constituent quark
model. The form factor calculated in the familiar light-front frame for
the $\pi$ meson [6] is reliable in this sense.
\par
For the spin-${1 \over 2}$ baryons the decomposition of the matrix elements
lead to the additional form factor $\alpha_2$ determined by Eqs.(4.4c)
or (4.4d), and $\alpha_1$ by (4.4a) or (4.4b). For the nucleon and $\Lambda$
hyperon Eqs.(4.4c) and (4.4d) give the same $\alpha_2$, Eqs. (4.4a) and (4.4b)
give the same $\alpha_1$. To see this we have developed analytical procedures
to calculate the current matrix elements. For
the nucleon and $\Lambda$ hyperon, we find that all $J^+$-current matrix
obey $<\lambda'=\uparrow\mid J^+\mid \lambda=\uparrow>
=<\lambda'=\downarrow\mid J^+\mid \lambda=\downarrow>$, while
$<\lambda'=\uparrow\mid J^+\mid \lambda=\downarrow>
=-<\lambda'=\downarrow\mid J^+\mid \lambda=\uparrow>$; and they
are real and $\omega$-independent. Thus, Eqs.(4.4a)-(4.4d) give consistent
and $\omega$-independent form factors $\alpha_2$ and $\alpha_1$. We
conclude for the nucleon and the $\Lambda$ hyperon that the form factors $F_1$
and $F_2$ calculated in the constituent quark model are
$\omega$-independent and not influenced by the nonphysical form factors
$\alpha_3$, $\alpha_4$ and $\alpha_5$.
\par
The free current matrix elements thus give the physical form factors in the
impulse approximation without the interference of the nonphysical form factors
for the spin-0 meson, the nucleon and the $\Lambda$ hyperon.  We expect the
N to $\Delta$ baryon transition matrix elements to have problems brought about
by the spin 3/2 of the $\Delta_{3,3}$.
\par
After this paper was submitted for publication we became aware of ref.24 where
the axial-vector matrix element of the pion is found insensitive to changes of
the light-cone direction in numerical studies.
\vskip 0.2in
\begin{center} {\bf          Acknowledgments   }\end{center}
\vskip 0.1in
X. Xu thanks the Department of Physics of the University of
Virginia for their hospitality. This work was supported in part by the U.S.
National Science Foundation.
\vfill\eject\par
\begin{center} {\bf       Appendix 1}\end{center}
\vskip 0.1in
In Eq.(2.7) the Euler angles $\alpha$ and $\beta$ determine $\vec n$. The
two axes $s_\mu^1$ and $s_\mu^2$ are fixed by the three Euler angles through
the rotation matrix in Eq.(2.3). For an investigation of the
$\omega$-dependence of the electromagnetic form factors, $s_\mu^1$ and
$s_\mu^2$ perpendicular to $\omega_\mu$ are irrelevant. Therefore, the angle
$\gamma$ is chosen to be the zero so that the $s_\mu^1$ and $s_\mu^2$ have the
simple form
$$
s_\mu^1=(0,-{n_xn_z \over d_1},-{n_yn_z \over d_1},d_1),~~~~~
s_\mu^2=(0,{n_y \over d_1}, -{n_x \over d_1}, 0)     \eqno (A1)
$$
with $d_1=-\sqrt{1-n_z^2}$.
The spin-up and spin-down spinors in Eqs. (2.12) are rewritten as
$$
u_{LF\uparrow}(p_{LF})={1 \over 2} \sqrt{1-n_z \over 2mp^+_{LF}}
\pmatrix{b_{LF}+{d_1 \over 1-n_z}p^R_{LF}       \cr
         -{n^R \over 1-n_z}b_{LF} + {n^R \over d_1} p^R_{LF}     \cr
         d_{LF} - {d_1 \over 1-n_z} p^R_{LF}        \cr
         -{n^R \over 1-n_z}d_{LF} + {n^R \over d_1} p^R_{LF},    \cr}
                       \eqno (A1)
$$
$$
u_{LF\downarrow}(p_{LF})={1 \over 2} \sqrt{1-n_z \over 2mp^+_{LF}}
\pmatrix{{n^L \over 1-n_z}b_{LF} -{n^L \over d_1}p^L_{LF}     \cr
         b_{LF} + {d_1 \over 1-n_z} p^L_{LF}        \cr
         -{n^L \over 1-n_z}d_{LF} + {n^L \over d_1} p^L_{LF}    \cr
         -d_{LF} -{d_1 \over 1-n_z} p^L_{LF}     \cr}
                          \eqno (A2)
$$
with $b_{LF}=p^+_{LF}+m$ and $d_{LF}=p^+_{LF}-m$.
If $n_x=n_y=0, {n^R \over d_1}={n^L \over d_1}=1$ and $n_z=-1$, then
the spinors reduce to the standard light-front spinors with Eq.(2.14),
$$
u_{\uparrow}(p)={1 \over 2\sqrt{mp^+}}
\pmatrix{b \cr    p_R  \cr   d   \cr   p_R   \cr},~~~~~~~~~~
u_{\downarrow}(p)={1 \over 2\sqrt{mp^+}}
\pmatrix{-p_L   \cr   b   \cr    p_L   \cr   -d   \cr}     \eqno (A4)
$$
with $b=p^++m$ and $d=p^+-m$.
\par
To calculate symbolically the matrix elements of the
electromagnetic current between hadron states using REDUCE, matrix elements of
$\gamma_{LF}$ for a single spinor,
$\bar {u}_{LF\lambda_k}(p_k) \Gamma u_{LF\lambda_i}(p_i)$,
must be calculated first. Note here that the momentum $p_k (p_i)$ may be that
of a quark or a hadron. The matrix elements are exhibited
in Tables I and II with recourse to following abbreviations,
$$
m_{ki}=\sqrt{x_km_kx_im_i},~~~~~~
A^{\pm}_{ki}={1 \over 2m_{ki}}(x_km_i \pm x_im_k),~~~~~~~
X_{ki}={2p^+_{LF} \over 2m_{ki}}x_kx_i,
$$
$$
K^{R,L}_{ki}={1 \over 2m_{ki}}(x_kp^{R,L}_{LFi}-x_ip^{R,L}_{LFk}), ~~~~~~~
W^{R,L\pm}_{ki}={2 \over 2m_{ki}p^+_{LF}}(m_kp^{R,L}_{LFi} \pm
m_ip^{R,L}_{LFk}),
$$
$$
Y^{RL\pm}_{ki}={2 \over 2m_{ki}p^+_{LF}}(m_km_i \pm p^R_{LFk}p^L_{LFi}),~~~~~~~
Y^{LR\pm}_{ki}={2 \over 2m_{ki}p^+_{LF}}(m_km_i \pm p^L_{LFk}p^R_{LFi}).
$$
The entries in Tables I,II are in agreement with those given by
Konen and Weber [1], if $n_x=n_y=0$, $n_z=-1$, and ${n^R \over d_1}=
{n^L \over d_1}=1$.

\vfill\eject\par
\begin{center} {\bf  References }\end{center}
\vskip 0.15in
\begin{description}
\item {[1]}I. G. Aznauryan, A. S. Bagdasaryan and N. L. Ter-Isaakyan, Phys.
Lett.
\par
{\bf 112B} (1982) 393, Yad. Fiz. {\bf 36}(1982)1743 [Sov. J. Nucl.
Phys. {\bf 36} (1982)] 743; H.~J.~Weber, Ann. Phys.(N. Y.){\bf 177} (1987) 38;
W.~Konen and H.~J.~Weber, Phys. Rev. {\bf D41} (1990) 2201.
\item {[2]}S.~Capstick and B.~D.~Keister, Phys. Rev. {\bf D51} (1995) 3598.
\item {[3]}F.~Schlumpf, Phys. Rev. {\bf D47} (1993) 4114; {\bf D51} (1995)
2262;
S.~J.~Brodsky and F.~Schlumpf, Phys. Lett. {\bf B329} (1994) 111.
\item {[4]}H.~J.~Weber, Phys. Rev. {\bf C41} (1990) 2783; I.~G.~Aznauryan and
A.~S.~Bagdasaryan, Yad. Fiz. {\bf 41} (1985) 249 [Sov. J. Nucl. Phys. {\bf 41}
 (1985) 158].
\item {[5]}J.~Bienkowska, Z.~Dziembowski and H.~J.~Weber, Phys. Rev. Lett.
{\bf 59} (1987) 624, 1790; H.~J.~Weber, Ann. Phys. (N.Y.) {\bf 207} (1991)
417; I. ~G. ~Aznauryan, Z. Phys. {\bf A346} (1993) 297.
\item {[6]}Z.~Dziembowski, Phys. Rev. {\bf D37} (1988) 768, 778;
I.~G.~Aznauryan
and K.~A.~Oganessyan, Phys. Lett. {\bf B249} (1990) 309; C.-R.~Ji and
P.~L.~Chung and S.~R.~Cotanch, Phys. Rev. {\bf D45} (1992) 4214.
\item {[7]}Z.~Dziembowski and L.~Mankiewicz, Phys. Rev. Lett. {\bf 55} (1985)
1839.
\item {[8]}H.~J.~Weber, Phys. Lett. {\bf B218} (1989) 267.
\item {[9]}L.~A.~Kondratyuk and M.~V.~Terent'ev, Sov. J. Nucl. Phys.
   {\bf 31} (1980) 561; H.~J.~Melosh, Phys. Rev. {\bf D9} (1974) 1095.
\item {[10]}M.~A.~Shifman, A.~I.~Vainshtein and V.~I.~Zakharov, Nucl. Phys.
{\bf B147} (1979) 385, 448, 519; B. L. Ioffe, ibid. {\bf B188} (1981) 317;
B. L. Ioffe and A. V. Smilga, ibid. {\bf B216} (1983) 373, {\bf B232} (1984)
109.
\item{[11]}P.~A.~M.~Dirac, Rev. Mod. Phys. {\bf 21} (1949) 392.
\item{[12]}We use the units $c=1=\hbar$ and $\gamma$-matrix conventions of
J.~D.~Bjorken and S.~D.~Drell, {\it Relativistic Quantum Mechanics},
(McGraw-Hill, New York, 1964).
\item {[13]}L. Susskind, Phys. Rev. {\bf 165} (1968) 1535; J. B. Kogut and
D. E. Soper, Phys. Rev. {\bf D1} (1970) 2901.
\item {[14]}V.~A.~Karmanov and A.~V.~Smirnov, Nucl. Phys. {\bf A546} (1992)
691; {\bf A575} (1994) 520.
\item {[15]}V.~A.~Karmanov, ZhETF {\bf 71} (1976) 399 [Sov. Phys. JETP {\bf 44}
 (1976) 210]; ZhETF {\bf 75} (1978) 1187 [Sov. Phys. JETP {\bf 48} (1978) 598].
\item {[16]}M.~G.~Fuda, Ann. Phys. {\bf 197} (1990) 265; {\bf 231} (1994) 1;
Phys. Rev. {\bf D41} (1990) 534; {\bf D42} (1990) 2898; {\bf D44} (1991) 1880;
M.~G.~Fuda and Y.~Zhang, Phys. Rev. {\bf C51} (1995) 23.
\item{[17]}F.~M.~Lev, Ann. Phys.(N.Y.){\bf 237} (1995) 355.
\item {[18]} H.~Leutwyler and J.~Stern, Ann. Phys. (N.Y.) {\bf 112} (1978) 94.
\item {[19]} L. L. Frankfurt, T. Frederico and M. I. Strikman, Phys. Rev. {\bf
C48} (1993) 2182.
\item {[20]}B.D. Keister, Phys. Rev. {\bf D49} (1994) 1500.
\item {[21]}S.~J.~Brodsky, T.~Huang and G.~P.~Lepage, in {\it Quarks and
Nuclear Forces}, eds. D. Fries and B. Zeitnitz, Springer Tracts in Modern
Physics, Vol.100, (Springer, New York, 1982).
\item {[22]}V.~G.~Kadyshevsky, ZhETF {\bf 46} (1964) 542, 872 [Sov. Phys. JETP
{\bf 19} (1964) 443, 597].
\item {[23]}S.~D.~Drell and T.~M.~Yan, Phys. Rev. Lett. {\bf 24} (1970) 181.
\item {[24]}A. Szczepaniak, C.-R. Ji and S.R. Cotanch, Phys. Rev. {\bf D} (Oct.
1995).
\end{description}

\vskip 0.1in
\vfill\eject\par
$$\vbox{\offinterlineskip
\halign{&\vrule#&\strut\ #\ \cr
\multispan{13}\hfil\bf TABLE I\hfil\cr
\noalign{\medskip}
\noalign{\hrule}
height3pt&\omit&&\omit&&\omit&&\omit&\cr
&\hfil$\lambda_i$\hfil&&\hfil$\Gamma$\hfil&&\hfil$(p\!\!\!/_{LFk})
\Gamma u_{LF\lambda_i}$
\hfil&&\hfil$\bar{u}_{LF\lambda_k} \Gamma u_{LF\lambda_i}$\hfil&\cr
height3pt&\omit&&\omit&&\omit&&\omit&\cr
\noalign{\hrule}
height3pt&\omit&&\omit&&\omit&&\omit&\cr
&$\uparrow$&&1&&$u_{LFk\uparrow}A^+_{ki}+u_{LFk\downarrow} {n^R \over d_1}
 K^R_{ki}$&
&$\delta_{\lambda_k,\uparrow} A^+_{ki}+\delta_{\lambda_k, \downarrow}
{n^R \over d_1} K^R_{ki}$&\cr

&$\downarrow$&&1&&$-u_{LFk\uparrow}{n^L \over d_1}K^L_{ki}+u_{LFk\downarrow}
A^+_{ki}$&&$
-\delta_{\lambda_k, \uparrow} {n^L \over d_1} K^L_{ki}+\delta_{\lambda_k,
\downarrow}A^+_{ki}$&\cr

&$\uparrow$&&$\gamma^+_{LF}$&&$u_{LFk\uparrow}X_{ki}$&
&$\delta_{\lambda_k,\uparrow}X_{ki}$&\cr

&$\downarrow$&&$\gamma^+_{LF}$&&$u_{LFk\downarrow}X_{ki}$&&$
\delta_{\lambda_k,\downarrow}X_{ki}$&\cr

&$\uparrow$&&$\gamma^-_{LF}$&&$u_{LFk\uparrow}Y^{LR+}_{ki}+u_{LFk\downarrow}
{n^R \over d_1}W^{R-}_{ki}$&&$\delta_{\lambda_k,\uparrow}Y^{LR+}_{ki}
+\delta_{\lambda_k,\downarrow}{n^R \over d_1} W^{R-}_{ki}$&\cr

&$\downarrow$&&$\gamma^-_{LF}$&&$-u_{LFk\uparrow}{n^L \over d_1}W^{L-}_{ki}
+u_{LFk\downarrow}
Y^{RL+}_{ki}$&&$-\delta_{\lambda_k,\uparrow}{n^L \over d_1}W^{L-}_{ki}+\delta_
{\lambda_k,\downarrow}Y^{RL+}_{ki}$&\cr

&$\uparrow$&&$\gamma^R_{LF}$&&$u_{LFk\uparrow}{x_kp^R_{LFi} \over m_{ki}}
$&&$\delta_{\lambda_k,\uparrow}{x_kp^R_{LFi} \over m_{ki}}$&\cr

&$\downarrow$&&$\gamma^R_{LF}$&&$2u_{LFk \uparrow} {n^L \over d_1} A^-_{ki}+
u_{LFk\downarrow}{x_ip^R_{LFk} \over m_{ki}}$&
&$2\delta_{\lambda_k, \uparrow} {n^L \over d_1} A^-_{ki} + \delta_
{\lambda_k,\downarrow} {x_ip^R_{LFk} \over m_{ki}}$&\cr

&$\uparrow$&&$\gamma^L_{LF}$&&$u_{LFk\uparrow}{x_ip^L_{LFk} \over m_{ki}}
-2u_{LFk \downarrow} {n^R \over d_1} A^-_{ki}$&
&$\delta_{\lambda_k,\uparrow}{x_ip^L_{LFk} \over m_{ki}}
-2\delta_{\lambda_k,\downarrow} {n^R \over d_1} A^-_{ki}$&\cr

&$\downarrow$&&$\gamma^L_{LF}$&&$u_{LFk\downarrow}
{x_kp^L_{LFi} \over m_{ki}}$&&$
\delta_{\lambda_k,\downarrow}{x_kp^L_{LFi} \over m_{ki}}$&\cr
height3pt&\omit&&\omit&&\omit&&\omit&\cr
\noalign{\hrule}\noalign{\medskip}
\multispan3\hfil&\multispan{10}\ {~~~~~~~~~~~~~~~~~~~~~~~~~~~~~~~~~~~~~~~~~~~~~
{}~~~~~~~~~~~~~~~~~~~~~~~~~~~~~~~~~~~~~~~~~~}\hfil\cr}}$$

\vfill\eject\par
$$\vbox{\offinterlineskip
\halign{&\vrule#&\strut\ #\ \cr
\multispan{13}\hfil\bf TABLE II\hfil\cr
\noalign{\medskip}
\noalign{\hrule}
height3pt&\omit&&\omit&&\omit&&\omit&\cr
&\hfil$\lambda_i$\hfil&&\hfil$\Gamma$\hfil&&\hfil$(p\!\!\!/_{LFk})
\Gamma u_{LF\lambda_i}$
\hfil&&\hfil$\bar{u}_{LF\lambda_k} \Gamma u_{LF\lambda_i}$\hfil&\cr
height3pt&\omit&&\omit&&\omit&&\omit&\cr
\noalign{\hrule}
height3pt&\omit&&\omit&&\omit&&\omit&\cr

&$\uparrow$&&$\gamma_5$&&$-u_{LFk \uparrow}A^-_{ki} +u_{LFk\downarrow}
{n^R \over d_1}
K^R_{ki}$&&$-\delta_{\lambda_k, \uparrow}A^-_{ki}+\delta_{\lambda_k,\downarrow}
{n^R \over d_1} K^R_{ki}$&\cr

&$\downarrow$&&$\gamma_5$&&$u_{LFk\uparrow}{n^L \over d_1} K^L_{ki}
+u_{LFk\downarrow}
A^-_{ki}$&&$\delta_{\lambda_k, \uparrow}{n^L \over d_1} K^L_{ki}
+\delta_{\lambda_k,\downarrow}A^-_{ki}$&\cr

&$\uparrow$&&$\gamma_5 \gamma^+_{LF}$&&$-u_{LFk\uparrow}X_{ki}$&
&$-\delta_{\lambda_k,\uparrow}X_{ki}$&\cr

&$\downarrow$&&$\gamma_5 \gamma^+_{LF}$&&$u_{LFk\downarrow}X_{ki}$&&$
\delta_{\lambda_k,\downarrow}X_{ki}$&\cr

&$\uparrow$&&$\gamma_5 \gamma^-_{LF}$&&$u_{LFk\uparrow}Y^{LR-}_{ki}
-u_{LFk\downarrow}
{n^R \over d_1}W^{R+}_{ki}$&&$\delta_{\lambda_k,\uparrow}Y^{LR-}_{ki}
-\delta_{\lambda_k,\downarrow}{n^R \over d_1} W^{R+}_{ki}$&\cr

&$\downarrow$&&$\gamma_5 \gamma^-_{LF}$&&$-u_{LFk\uparrow}{n^L \over d_1}
W^{L+}_{ki}-u_{LFk\downarrow}
Y^{RL-}_{ki}$&&$-\delta_{\lambda_k,\uparrow}{n^L \over d_1}W^{L+}_{ki}-\delta_
{\lambda_k,\downarrow}Y^{RL-}_{ki}$&\cr

&$\uparrow$&&$\gamma_5 \gamma^R_{LF}$&&$-u_{LFk\uparrow}{x_kp^R_{LFi} \over
m_{ki}}
$&&$-\delta_{\lambda_k,\uparrow}{x_kp^R_{LFi} \over m_{ki}}$&\cr

&$\downarrow$&&$\gamma_5 \gamma^R_{LF}$&&$-2u_{LFk \uparrow} {n^L \over d_1}
A^+_{ki}+
u_{LFk\downarrow}{x_ip^R_{LFk} \over m_{ki}}$&
&$-2\delta_{\lambda_k, \uparrow} {n^L \over d_1} A^+_{ki} + \delta_
{\lambda_k,\downarrow} {x_ip^R_{LFk} \over m_{ki}}$&\cr

&$\uparrow$&&$\gamma_5 \gamma^L_{LF}$&&$-u_{LFk\uparrow}{x_ip^L_{LFk} \over
m_{ki}}
-2u_{LFk \downarrow} {n^R \over d_1} A^+_{ki}$&
&$-\delta_{\lambda_k,\uparrow}{x_ip^L_{LFk} \over m_{ki}}
-2\delta_{\lambda_k,\downarrow} {n^R \over d_1} A^+_{ki}$&\cr

&$\downarrow$&&$\gamma_5 \gamma^L_{LF}$&&$u_{LFk\downarrow}
{x_kp^L_{LFi} \over m_{ki}}$&&$
\delta_{\lambda_k,\downarrow}{x_kp^L_{LFi} \over m_{ki}}$&\cr
height3pt&\omit&&\omit&&\omit&&\omit&\cr
\noalign{\hrule}\noalign{\medskip}
\multispan3\hfil&\multispan{10}\ {~~~~~~~~~~~~~~~~~~~~~~~~~~~~~~~~~~~~~~~~~~~~~
{}~~~~~~~~~~~~~~~~~~~~~~~~~~~~~~~~~~~~~~~~~~~~~}\hfil\cr}}$$
\end{document}